\documentstyle[12pt,epsf]{article}
\def\beq{\begin{equation}}
\def\eeq{\end{equation}}
\def\noi{\noindent}
\begin{document}
\vbox to 1 truecm {}
\begin{center}{\bf JETS IN PHOTON-PHOTON COLLISIONS} \\
\vbox to 1.5 truecm {}
M. FONTANNAZ\\
{\it Laboratoire de Physique Th\'eorique et Hautes Energies$^{\ast}$}\\
{\it Universit\'e de Paris XI, b\^atiment 211, 91405 Orsay Cedex, France}\\
\end{center}
\vspace{2 cm}

\noindent {\bf Abstract} \par
We study jet production in photon-photon reactions at the next-to-leading
logarithm
accuracy. The discussion of the theoretical uncertainties and the
role of the quark and gluon distributions in the photon is emphasized. The
phenomenology at TRISTAN
energies is discussed and predictions are made for LEP 200. \\

\vbox to 2 truecm {}

\noindent {\it Invited talk at the ``Two-Photon Physics from DAPHNE To LEP 200
and Beyond'' Meeting,
Paris, February 1994} \par
\vbox to 3 truecm {}
\noindent LPTHE Orsay 94/44 \\
\noindent  May 1994 \\
\vspace{2 cm}
\vskip 1 mm \hrule \vskip 1mm \noi{$^*$Laboratoire
associ\'e au Centre National de la Recherche Scientifique - URA 63}

\newpage
\begin{center}{\bf JETS IN PHOTON-PHOTON COLLISIONS} \\
\vbox to 1.5 truecm {}
M. FONTANNAZ\\
{\it Laboratoire de Physique Th\'eorique et Hautes Energies$^{\ast}$}\\
{\it Universit\'e de Paris XI, b\^atiment 211, 91405 Orsay Cedex, France}\\
\end{center}
\vspace{5 mm}
\begin{quote}
\begin{center}{\tenrm ABSTRACT} \end{center}
\ \ \ {\tenrm We study jet production in photon-photon reactions at the
next-to-leading lo\-ga\-rithm
accuracy. The discussion of the theoretical uncertainties and the
role of the quark and gluon distributions in the photon is emphasized. The
phenomenology at TRISTAN
energies is discussed and predictions are made for LEP 200.}
\end{quote}
\vspace{5 mm}
\noi {\bf 1. Introduction} \par
\vspace{5 mm}
Very recently, several experimental collaborations have published results
concerning large
transverse momenta phenomena in $\gamma - \gamma$ collisions where the incoming
photons are real or
quasi-real. In particular, the AMY$^{1}$ and TOPAZ$^{2}$ collaborations have
measured the high
transverse momentum jet rates at $\sqrt{s}_{e^+e^-} = 58 \ {\rm GeV}$. At a
lower energy
($\sqrt{s}_{e^+e^-} = 29 \ {\rm GeV}$) a related observable, namely the
large-$p_T$ particle
spectrum, has been measured at the MARK II$^{3}$ detector. Theoretical work on
this subject started
long ago$^{4,5}$. Here we discuss the QCD predictions, at the next-to-leading
logarithm accuracy, of
the production of jets in $\gamma - \gamma$ collisions$^{6}$. This study seems
appropriate at this
time since the above data are rather precise and, furthermore, data at higher
energies are coming
from LEP$^{7}$. It is well known that the rate of $\gamma - \gamma$ events
increases with energy in
contrast to the rate of $e^+e^-$ annihilation into a virtual photon in the
$s$-channel and therefore
$\gamma - \gamma$ large-$p_T$ phenomena will provide an increasingly important
background in the
search of new phenomena at very high energies.

The production of large-$p_T$ jets is schematically represented in Fig. 1, in
which we see that a
photon can take part in the reaction in two different ways. Either it directly
couples to a quark
of the subprocess (as the photon of momentum $p'$, $|p'^2| << \Lambda^2$) or it
interacts through
its quark and gluon contents (as the photon of momentum $p$, $|p^2| <<
\Lambda^2$). With ${\cal
P}_{\gamma}^i$ the distributions of parton in the photon, we symbolically write
the inclusive jet
cross-section as
\beq
{d\sigma^{jet} \over d \vec{p}_{\bot} \ d \eta} = \sum_i {\cal P}_{\gamma}^i
\otimes
\widehat{\sigma} \left ( \gamma i \to \hbox{jet} \right )
 \eeq
\vskip 1 mm \hrule \vskip 1mm \noi{\tenrm $^*$Laboratoire
associ\'e au Centre National de la Recherche Scientifique - URA 63}
\newpage
$$
\epsfbox{desco1}
$$
\noindent where $\widehat{\sigma}$ is the subprocess cross-section ($\eta$ is
the jet
pseudo-rapidity). Expanding ${\cal P}_{\gamma}^i$ and $\widehat{\sigma}$ in
powers of $\alpha_s$,
we obtain an expression

\beq
{d \sigma^{jet} \over d \vec{p}_{\bot} \ d \eta} = \sum_i \left ( {4 \pi \over
\alpha_s(p_{\bot}^2)} a_i + b_i \right ) \otimes \left ( \alpha \
\alpha_s(p_{\bot}^2)
\widehat{\sigma}_i^{BORN} + \alpha \ \alpha_s^2(p_{\bot}^2) K_i \right )
\eeq

\noindent which shows the Leading Logarithm (LL) contributions to the
cross-section (associated with
$a_i$ and $\widehat{\sigma}_i^{BORN}$), and the Higher Order (HO) QCD
corrections coming from $b_i$
and $K_i$.

The term $b_i$ describes the effects of the HO corrections to the evolution
equations of the quark
and gluon distributions in the photon ; it is discussed in the next section in
which we also
address the issue of the non perturbative part of these distributions. As we
shall see below, the
contributions to $d\sigma^{jet}/d\vec{p}_{\bot} \ d \eta$ involving the
distributions ${\cal
P}_{\gamma}^i$ are large, and it is worth spending a moment in discussing all
the peculiarities of
the latter. Studied by Witten$^{8}$ a long time ago, they are hard to measure ;
the quark
contents of the photon can be obtained from the Deep Inelastic Scattering of a
real photon or a
virtual photon $\gamma^{\ast}$, but the gluon is more difficult to observe
directly. Only the
recent photoproduction experiments at HERA$^{9,10,11}$, and the $\gamma \gamma$
collisions at
TRISTAN$^{1,2}$ give a privileged access to this distribution.

In section 3 we will discuss the HO corrections $K_i$ to the subprocess and
show how they make the
inclusive jet cross-section more stable with respect to variations of the
factorization scale.
This fact makes possible a quantitative comparison between theoretical
predictions and data.
\vskip 5 mm
\noi {\bf 2. The Photon structure function}
\vskip 5 mm
The parton contents of the photon$^{12}$ can be measured in deep inelastic
scat\-te\-ring experiments
in which the virtual photon $\gamma^{\ast}$ of momentum $q$ ($Q^2 = - q^2 >>
\Lambda^2$) probes the
short distance behavior of the real photon $\gamma$ of momentum $p$. The
structure function
$F_2^{\gamma}$ of this reaction is proportional, in the LL approximation, to
the quark distributions
in the real photon

\beq
F_2^{\gamma}(x, Q^2) = x \sum_{f=1}^{n_f} e_f^2 \left ( q_{\gamma}^f(x, Q^2) +
\bar{q}_{\gamma}^f(x,
Q^2) \right ) \ \ \ .
\eeq

\noi The sum in (3) run over the quark flavors and $x = Q^2/2p.q$.

It is instructive to consider the contribution to $F_2^{\gamma}$ of the lowest
order diagrams of
Fig. 2. Contrarily to the case of a hadronic target, the lower part of the
diagram is known~: it is
given by the coupling of photon to quark.
$$
\epsfbox{desco2}
$$
\noindent This contribution is therefore exactly calculable, with the following
result for a quark of
charge $e_f$~:

\beq
F_2^{\gamma}(x, Q^2) = 3e_f^4 {\alpha \over \pi} x \left \{ \left ( x^2 +
(1-x)^2 \right )
\ell n {Q^2 \over m_f^2} + \left ( x^2 + (1-x)^2 \right ) \ell n {1-x \over x}
+ 8x(1-x) -1
\right \} .
\eeq

However our result (4) is not directly related to a physical process, because
it depends on the
unknown quark mass $m_f$, used as a cut-off to regularize a logarithmic
divergence. Actually this
perturbative approach is certainly not valid when the virtuality $k^2$ of the
exchanged quark
becomes small. We then go into a non perturbative domain where we lack
theoretical tools and we must
resort to models to describe non perturbative (NP) contributions to
$F_2^{\gamma}$. A popular model
is the ``Vector Meson Do\-mi\-nan\-ce Model'' (VDM) which consider that the
real photon couple to
vector mesons. Therefore the real photon, besides a direct coupling to a $q
\bar{q}$ pair (Fig. 3.a),
has a VDM component (Fig. 3.b)
$$
\epsfbox {desco3}
$$
\noi which is also probed by the virtual photon.

The latter component contributes to $F_2^{\gamma}$ and must be added to
expression (4). Keeping only
the term in (4) proportional to $Log \ Q^2/m_f^2$ (LL approximation), we write

\beq
F_2^{\gamma}(x, Q^2) = 3 e_f^4 {\alpha \over \pi} x \left ( x^2 + (1 - x)^2
\right ) \ell n {Q^2 \over
Q_0^2} + x \sum_{V=\rho, \omega , \phi} e_f^2 \left ( q_f^V(x) + \bar{q}_f^V(x)
\right ) \ \ \ .
\eeq

\noi The scale $Q_0^2$ is the value of $Q^2$ at which the perturbative approach
is no more valid.
The perturbative contribution vanishes and $F_2^{\gamma}$ is described only by
the non perturbative
contribution $q_f^{NP}(x) = q_f^V(x)$ which describes the quark contents of
vector mesons.

We have to keep in mind that this way of treating the non perturbative part of
$F_2^{\gamma}$ is
due to our lack of theoretical understanding of this contribution. There are
other
approaches$^{13}$, especially that of ref. 14 which takes into account the
interaction between the
quarks and the gluon condensate. These different approaches must ultimately be
compared with
experiment.

QCD corrections to the diagrams of Fig. 2 do not change the basic structure of
expression (5). In
the LL approximation$^{8, 15}$, the perturbative quark distribution is given by
the sum of ladder
diagrams (Fig. 4) (for simplicity we forget the gluons and consider only the
non singlet quark
distribution)
$$
\epsfbox{desco4}
$$
\begin{center}
{\tenrm Fig. 4 : Ladder diagram contribution to $F_2^{\gamma}$ (the thin line
cuts final
partons)}
\end{center}

\noi and the non perturbative part acquires a $Q^2$-dependence which is
identical to that
of a quark distribution in a hadron. The sum of the ladder diagrams can be
written in a very compact
form (AN is for anomalous, a designation introduced in ref. 8)

\beq
q_{\gamma}^{AN}(n, Q^2) = {\alpha \over 2 \pi}
\int_{\alpha_s(Q_0^2)}^{\alpha_s(Q^2)} {d \alpha '_s
\over \beta (\alpha '_s)} k^{(0)}(n) e^{\int_{\alpha '_s}^{\alpha_s(Q^2)} {d
\alpha ''_s \over
\beta(\alpha ''_s)} {\alpha ''_s \over 2 \pi} P^{(0)}(n)}
\eeq

\noi in terms of moments of the Altarelli-Parisi kernels $k^{(0)}(n) = \int_0^1
dx \ x^{n-1}
k^{(0)}(x)$ and $P^{(0)}(n)$ describing the splitting of a photon into a $q
\bar{q}$ pair (the
bottom rung of the ladder) and the splitting of a quark into a quark and a
gluon (the other
rungs of the ladder). The beta function has the usual definition $\partial
\alpha_s(Q^2)/\partial \ell n \ Q^2 = \beta (\alpha_s) = - \alpha_s ({\alpha_s
\over 4 \pi} \beta_0
+ ({\alpha_s \over 4 \pi})^2 \beta_1 + \cdots )$. The total quark distribution
is given by

\beq
q_{\gamma}(n, Q^2) = q_{\gamma}^{AN}(n, Q^2) + q_{\gamma}^{NP}(n, Q^2)
\eeq

\noi which verifies the inhomogeneous equation

\beq
Q^2 {\partial q_{\gamma}(n, Q^2) \over \partial Q^2} = {\alpha \over 2 \pi}
k^{(0)}(n) +
{\alpha_s(Q^2) \over 2 \pi} P^{(0)}(n) q_{\gamma}(n, Q^2) \ \ \ .
\eeq

\noi As in (5) we introduce the boundary condition$^{16}$ $Q_0^2$ in (6) so
that
$q_{\gamma}^{AN}(n, Q_0^2)$ vanishes when $Q^2 = Q_0^2$.

The modifications of these LL results due to HO QCD corrections$^{17, 18, 19,
20, 21, 22}$ are
obtained by replacing the LL kernels of (6) by kernels involving HO
contributions

\begin{eqnarray}
k(n) & = & {\alpha \over 2 \pi} k^{(0)}(n) + {\alpha \over 2 \pi} {\alpha_s
\over 2 \pi} k^{(1)}(n)
+ \cdots \nonumber \\
P(n) & = & {\alpha_s \over 2 \pi} P^{(0)}(n) + \left ( {\alpha_s \over 2 \pi}
\right )^2 P^{(1)}(n)
+ \cdots \ \ \ ,
\end{eqnarray}

\noi and by a modification of the expression of $F_2^{\gamma}$ in terms of
parton distributions (the
gluon contribution is now explicitly written)

\beq
F_2^{\gamma}(n - 1, Q^2) = e_f^2 C_q(n, Q^2) \left ( q_{\gamma}(n, Q^2) +
\bar{q}_{\gamma}(n, Q^2)
\right ) + C_g(n, Q^2) g_{\gamma}(n, Q^2) + C_{\gamma}(n)
\eeq

\noi where $C_{\gamma}$ is the ``direct term'', given by the part of (4) not
proportional to $\ell n
{Q^2 \over m_f^2}$. $C_q$ and $C_g$ are the well-known Wilson coefficients
which are identical to
those found in the case of a hadronic target.

The expression of $q_{\gamma}^{AN}$ including the HO QCD corrections is
obtained from (6) with the
kernels (9). It can be written as an expansion in $\alpha_s$

\beq
q_{\gamma}^{AN}(n, Q^2) = {4 \pi \over \alpha_s(Q^2)} \left [ 1 - r_0^{1-d}
\right ] \left ( a +
\alpha_s(Q^2)c \right ) + \left [ 1 - r_0^{-d} \right ] b + 0 \left ( \alpha_s
(Q^2) \right )  \eeq

\noi where $d = 2 P^{(0)}(n)/\beta_0$ and $r_0 =
\alpha_s(Q^2)/\alpha_s(Q_0^2)$. The higher order
corrections are contained in the terms $b$ and $c$ which are functions of
$k^{(1)}(n)$, $P^{(1)}(n)$
and $\beta_1$. They are required to perform a consistent beyond LL calculation
of the jet
cross-section.

We can now compare data$^{23}$ with the theoretical predictions$^{21}$. The
latter are obtained in
the $\overline{MS}$ factorization scheme with $\Lambda_{\overline{MS}} = 200 \
{\rm MeV}$ ; the non
perturbative contribution is described by VDM with $Q_0^2 = .5 \ {\rm GeV}
\simeq m_f^2$. We get a
reasonable agreement between theory and experiment as can be seen from Fig. 5.
More comparisons
can be found in ref. 21. From this study we conclude that the quark
distribution in the photon is
constrained by data on $F_2^{\gamma}$, but that they are not accurate enough to
teach us something
about the non perturbative contribution and the value of $Q_0^2$.

\vbox to 8 cm {}
\begin{center}
{\tenrm Fig. 5 : Comparison of JADE data with theoretical predictions. The
dashed curve corresponds
to a non perturbative (VDM) input set equal to zero.}
\end{center}

A delicate point when working beyond the LL approximation is that of the
factorization scheme. A
change in the factorization scheme is translated into a change in $k^{(1)}$ and
$C_{\gamma}$
but in such a way that the physical quantity $F_2^{\gamma}$ remains unmodified
(at order
$\alpha_s^0$). On the other hand $q^{\gamma}$ is not an invariant with respect
to the factorization
scheme and a change in $k^{(1)}$ causes modifications in $q^{AN}$
\underbar{and} $q_{\gamma}^{NP}$.
Therefore the separation (7) in a perturbative and a non perturbative part is
\underbar{not}
scheme invariant and the statement that $q_{\gamma}^{NP}$ can be described by
VDM has no meaning,
unless one specifies in which factorization scheme it is valid. This part is
discussed in details
in ref. 21.

\vskip 5 mm
\noi {\bf 3. Jets in photon-photon collisions}
\vskip 5 mm
The jet production in photon-photon collisions gives a privileged access to
parton distributions
in the photon~; contributions to the cross-section due to these distributions
are indeed
important as we shall see below. Fig. 6 shows the various processes
contributing at order
$\alpha^2$~: the direct term (Fig. 6.a), the terms in which a photon interacts
via its parton
component (Fig. 6.b) and terms in which the two photons interact via their
parton contents. Although
the various subprocesses are not of the same order in $\alpha_s$, the final
result is of order
$\alpha^2$ because of the peculiar $\alpha_s$-behaviour of the parton
distributions. Fig. 6 only
shows the Born terms of the various subprocesses. HO QCD corrections to these
Born terms are
available from the literature$^{24}$ at the expense of a few modifications of
the color factors. They

\vbox to 10 cm {}

\noi give rise, for instance, to terms $K_i$ in expression (2) which describes
the situation in which
a photon interacts via its parton component. As we can see from this example,
the beyond LL
corrections calculated so far allow us to make predictions at order $\alpha^2
\alpha_s$ for the
one-jet inclusive cross-section $d \sigma^{jet}/d\vec{p}_{\bot} \ d \eta$. We
used everywhere the
$\overline{MS}$ factorization and renormalization scheme in calculating the HO
corrections $K$ to the
Born subprocesses, and we consistently use parton distributions in the photon
in the $\overline{MS}$
scheme. We saw in the preceding section that the non perturbative part of these
distributions is not
well determined. Therefore we have an arbitrary normalization factor in front
of the VDM component in
order to check the sensitivity of the cross-section to this non perturbative
input, and to explore
whether it is possible to constrain it from data.

Before discussing various consequences of these HO corrections to the
cross-section, we still
have to perform a convolution between the $\gamma \gamma \to $ jet
cross-section considered
until now and the photon distribution in the electron, in order to obtain the
cross-section of
the reaction $e^+e^- \to \hbox{jet} + X$ which is the one experimentally
observed. We describe
the photon contents of the electron by the Weizs\"acker-Williams approximation,
taking into
account the anti-tag conditions of TOPAZ$^{2}$

\beq
F_{\gamma /e}(z, E) = {\alpha \over 2 \pi z} \left ( 1 + (1 - z)^2 \right )
\int_{Q^2_{min}}^{Q^2_{max}} {dQ^2 \over Q^2}
\eeq

\noi with $Q^2_{min} = m_e^2 z^2/(1 - z)$ and $Q_{max}^2 = E^2 \theta_{max}^2(1
- z)$~; $E = 29 \ {\rm
GeV}$ is the beam energy and $\theta_{max} = 3.2^o$ defines the anti-tag
condition. This way of
treating the photon as real and collinear amounts to neglect the effects of the
virtuality and
of the transverse momentum of the photon in the calculation of the
cross-section. These effects
are studied in details in ref. 6 and correcting factors are introduced to take
them into
account.

It is well-known that HO QCD corrections make the inclusive cross-sections more
stable with
respect to variations of the factorization and renormalization scales which
appear in the
perturbative calculations. This fact is illustrated in Fig. 7 which shows the
variations of the
jet cross-section as a function of the factorization scale $M$ and
renormalization scale
$\mu$. We see that we get a much flatter surface in a large domain in $\mu$ and
$M$ (Fig. 7.b)
when HO corrections are added to the LL cross-section shown in Fig. 7.a. For
the comparison
with TOPAZ data, we use $\mu = M = p_{\bot}$.

\vbox to 9 cm {}
\begin{center}
{\tenrm Fig. 7 : Variation of the cross section ${d \sigma^{e^+e^- \to
\hbox{jet}} \over d
\vec{p}_T \ d \eta}$ at $\sqrt{s}_{e^+e^-} = 58 \ {\rm GeV}$, $p_T = 5.24 \
{\rm GeV/c}$, $\eta = 0$,
with the factorization scale $M$ and the renormalization scale $\mu$ : a)
leading logarithm
predictions ; b) next-to-leading logarithm predictions. Note that in order to
fully display the shape
of the surface we rotated the figure by $90^o$ compared to fig. a.}
\end{center}

This comparison is shown in Fig. 8 where one notices a nice agreement between
theory and
experiment. For values of $p_{\bot}$ smaller than 5 GeV/c, the dotted curve is
above the
data. This is due to the fact that the present theoretical calculation neglects
the charm
quark mass and overestimates the cross-section. This effect is negligible for
$p_{\bot}$
larger than 5 GeV/c and the three theoretical curves corresponding to three
different non
perturbative inputs are in agreement with data. This result shows the weak
sensitivity of the
predictions to the non perturbative input (taking as reference scale the
accuracy of the
data). On the other hand we see that the direct term contributions is a factor
3 (at
$p_{\bot} = 5 \ {\rm GeV/c}$) below the data, a fact which clearly demonstrates
the
sensitivity of the jet cross-section to the parton component of the photon. Let
us finally
notice that the gluon contents of the photon plays an important role,
contributing to more
than 30 $\%$ of the total cross-section (at $p_{\bot} = 5 \ {\rm GeV/c}$).

In Fig. 9 we show the predictions for the jet $p_{\bot}$-spectrum at LEP 200
with a no-tag
condition. Despite the large scales involved it is interesting to remark that
the
non-perturbative term still gives a large contribution up to rather high
$p_{\bot}$ values.
This is related to the fact that higher energy experiments tend to probe
smaller $x$ values of
the photon structure function where the hadronic component is important.

\vbox to 9 cm {}
\noi{\tenrm Fig. 8 : TOPAZ data on inclusive jet production \hspace{5 mm} Fig.
9 : Single
jet production, via the}\\
\noi{\tenrm and theoretical predictions for $\int_{.7}^{.7} d \eta {d
\sigma^{e^+e^- \to
\hbox{jet}} \over dp_T \ d \eta}$. \hskip 5 mm two-photon
mechanism at LEP 200 for a} \\
\noi{\tenrm The top curve is the theoretical prediction based \hskip 7 mm
no-tag
experiment and $\eta = 0$. The curves} \\
\noi{\tenrm  on the standard photon structure functions, the \hskip 9 mm have
the same meaning as in Fig. 8.}
\\  \noi{\tenrm middle one is based on structure functions with} \\
\noi{\tenrm half the VDM input, and the lower one is based} \\
\noi{\tenrm on the perturbative component only. The dash-} \\
\noi{\tenrm dotted curve is the ``direct contribution''.} \\

\vskip 5 mm
\noi{\bf References}
\vskip 5 mm
\begin{description}
\item{1.\ } AMY collaboration : R. Tanaka et al., Phys. Lett. {\bf B277} (1992)
215.
\item{2.\ } TOPAZ collaboration : H. Hayashii et al., Phys. Lett. {\bf B314}
(1993) 149.
\item{3.\ } MARK II collaboration : D. Cords et al., Phys. Lett. {\bf B302}
(1993) 341.
\item{4.\ } S. J. Brodsky, T. DeGrand, J. Gunion and J. Weis, Phys. Rev. {\bf
19} (1979) 1418 ; \\ K.
Kajantie and R. Raitio, Nucl. Phys. {\bf B159} (1979) 528 ; \\ S. P. Li and H.
C. Liu, Phys. Lett.
{\bf 143B} (1984) 489 ; \\ A. V. Efremov, S. V. Ivanov and G. P. Korchemskii,
Sov. Jour. Nuc. Phys.
{\bf 39} (1984) 987 ; \\ F. A. Berends, Z. Kunszt and R. Gastmans, Nucl. Phys.
{\bf B182} (1984) 987
; \\ P. Aurenche, R. Baier, A. Douiri, M. Fontannaz and D. Schiff, Z. Phys.
{\bf C29} (1985) 423.
\item{5.\ } M. Drees and R. M. Godbole, Nucl. Phys. {\bf B339} (1990) 355 ; \\
M. Drees, preprint DESY
92-065, May 1992, Proceedings of the IXth International Workshop on
Photon-Photon Collisions, La
Jolla, CA, USA.
\item{6.\ } P. Aurenche, J.-Ph. Guillet, M. Fontannaz, Y. Shimizu, J. Fujimoto,
K. Kato, KEK
preprint 93-180, LAPP-TH 436/93, LPTHE-Orsay 93/47.
\item{7.\ } ALEPH collaboration : D. Buskulic et
al., CERN-PPE-93-94, June 1993. \\ DELPHI collaboration : P. Abreu et al.,
CERN-PPE-94/04.
\item{8.\ } E. Witten, Nucl. Phys. {\bf B120} (1977) 189.   \item{9.\ } H1
collaboration : T. Ahmed
et al., Phys. Lett. {\bf B297} (1992) 205 ; T. Abt et al., Phys. Lett. {\bf
314} (1993) 436.
\item{10.\ } ZEUS collaboration : M. Derrick et al., Phys. Lett. {\bf B297}
(1992) 404.
\item{11.\ } For a review on ``The Photon Structure Function at HERA'', see M.
Fontannaz, Orsay
preprint LPTHE 93-22, talk given at the 21st International meeting on
Fundamental physics, Miraflores
de la Sierra, Spain, May 1993.
\item{12.\ } Reviews on the photon structure function may be found in \\
C. Berger and W. Wagner,
Phys. Rep. {\bf 146} (1987) 1 ; \\
H. Kolanski and P. Zerwas, in High Energy $e^+e^-$ physics, World Scientific,
Singapore, 1988,
Eds. A. Ali and P. S\"oding ; \\
J. H. Da Luz Vieira and J. K. Storrow, Z. Phys. {\bf C51} (1991) 241.
\item{13.\ } I. Antoniadis and G. Grunberg, Nucl. Phys. {\bf B213} (1983) 455.
\item{14.\ } A. S. Gorski, B. L. Ioffe, A. Yu Khodjaminian, A. Oganesian, Z.
Phys. {\bf C44}
(1989) 523.
\item{15.\ } A review on QCD corrections, and previous references on the
subject can be found in~:
F. M. Borzumati and G. A. Schuler, Z. Phys. {\bf C58} (1993) 139.
\item{16.\ } M. Gl\"uck and E. Reya, Phys. Rev. {\bf D28} (1983) 2749.
\item{17.\ } W. A. Bardeen and A. J. Buras, Phys. Rev. {\bf D20} (1979) 166 ;
{\bf 21} (1980) 2041
(E).
\item{18.\ } M. Fontannaz and E. Pilon, Phys. Rev. {\bf D45} (1992) 382.
\item{19.\ } L. E. Gordon and J. K. Storrow, Z. Phys. {\bf C56} (1992) 307.
\item{20.\ } M. Gl\"uck, E. Reya and A. Vogt, Phys. Rev. {\bf D45} (1992) 3986,
{\bf D46} (1992) 1973.
\item{21.\ } P. Aurenche, M. Fontannaz and J. Ph. Guillet, preprint
ENSLAPP-A-435-93, LPTHE Orsay
93-37.
\item{22.\ } E. Laenen, S. Riemersma, J. Smith and W. L. van Neerven,
University of Leiden
preprint ITP-SB-93-46.
\item{23.\ } JADE collaboration, W. Bartel et al., Z. Phys. {\bf C24} (1984)
231.
\item{24.\ } F. Aversa, P. Chiappetta, M. Greco and J. Ph.
Guillet, Nucl. Phys. {\bf B327} (1989) 105 ;  Z. Phys. {\bf C46} (1990) 401 ;
Phys. Rev. {\bf 65}
(1990) 401 ; \\ S. D. Ellis, D. Kunzst and D. E. Soper, Phys. Rev. {\bf D40}
(1989) 2188 ; Phys. Rev.
Lett. {\bf 64} (1990) 2121.  \end{description}

\end{document}